\documentclass[twocolumn,amsmath,amssymb,nofootinbib]{revtex4}

\usepackage{graphicx}
\usepackage{dcolumn}
\usepackage{bm}

\newcommand{\be}{\begin{equation}}
\newcommand{\ee}{\end{equation}}
\newcommand{\bea}{\begin{eqnarray}}
\newcommand{\eea}{\end{eqnarray}}
\newcommand{\ba}{\begin{eqnarray}}
\newcommand{\ea}{\end{eqnarray}}

\newcommand{\beq}{\begin{equation}}
\newcommand{\eeq}{\end{equation}}
\newcommand{\beqa}{\begin{eqnarray}}
\newcommand{\eeqa}{\end{eqnarray}}
\newcommand{\beqar}{\begin{eqnarray*}}
\newcommand{\eeqar}{\end{eqnarray*}}

\newcommand{\reef}[1]{(\ref{#1})}
\newcommand{\ssc}{\scriptscriptstyle}
\newcommand{\eg}{{\it e.g.,}\ }
\newcommand{\ie}{{\it i.e.,}\ }

\newcommand{\labell}[1]{\label{#1}} 

\newcommand{\lp}{\ell_{\mt P}}

\def\t6 {T_\mt{D6}}

\newcommand{\R}{L} 

\newcommand{\mt}[1]{\textrm{\tiny #1}}

\newcommand{\rhoa}{\rho_{\ssc A}}
\newcommand{\ha}{H_{\!\ssc A}}
\newcommand{\cA}{{\cal A}}
\newcommand{\m}{v}
\newcommand{\ren}{R\'enyi\ }

\usepackage[bookmarks=false]{hyperref} 
\hypersetup{pdfstartview=FitH,pdfhighlight=/O,colorlinks=false}

\begin{document}

\preprint{arXiv:1212.nnnn [hep-th]}

\title{On the Architecture of Spacetime Geometry}

\author{Eugenio Bianchi and Robert C. Myers}
 \affiliation{Perimeter Institute for Theoretical Physics, Waterloo,
Ontario N2L 2Y5, Canada }


\begin{abstract}
We propose entanglement entropy as a probe of the architecture of spacetime in
quantum gravity. We argue that the leading contribution to this entropy
satisfies an area law for any sufficiently large region in a smooth spacetime,
which, in fact, is given by the Bekenstein-Hawking formula. This conjecture is
supported by various lines of evidence from perturbative quantum gravity,
simplified models of induced gravity and loop quantum gravity, as well as the
AdS/CFT correspondence.

\end{abstract}

\maketitle

\noindent {\bf Introduction:} One of the most remarkable discoveries in
fundamental physics was the realization that black hole horizons carry entropy
\cite{BH:1973}. This entropy is manifest in the spacetime geometry, as
expressed by the Bekenstein-Hawking formula:
\be S_{BH} = \frac{k_B\,c^3}{\hbar}\ \frac{\cal A}{4G} \,, \labell{sbh} \ee
where $\cal A$ is the area of the horizon. 
In fact, this result is easily extended to any Killing horizon, including de
Sitter \cite{DS} and Rindler \cite{ray} horizons. Of course, the above result
applies for black hole solutions of Einstein's equations (in any number of
dimensions). This result can also be extended to account for higher curvature
interactions in the gravitational theory and the corresponding expression for
the horizon entropy, the `Wald entropy' \cite{WaldEnt}, again has a geometric
character. This celebrated formula \reef{sbh} draws an unexpected connection
between spacetime geometry, thermodynamics, quantum theory and gravity. In
fact, interest in this result has long been sparked by the possibility that it
provides a window into the nature of quantum gravity. Certainly, reproducing
eq.~\reef{sbh} in terms of a microscopic description is now regarded as a
necessary hallmark for a consistent theory of quantum gravity.

Of course, this expression \reef{sbh} also reminds us that the physical
constants in Nature can be combined to yield a fundamental length, the Planck
scale: $\lp^{d-2}=8\pi\,G\, \hbar/c^3$ in $d$ spacetime dimensions. Hence the
geometric entropy \reef{sbh} of the horizon is simply the horizon area measured
in units of the Planck scale:
\be S_{geom} = 2\pi\,\frac{{\cal A}}{\lp^{d-2}}\ +\ \cdots\,. \labell{prop0}
\ee
Above we have also set Boltzmann's constant to unity and indicated the
possibility of contributions subleading to the area term.

In the present paper, we propose that eq.~\reef{prop0} in fact has much wider
applicability with the following general conjecture:
 \vskip 3pt
\noindent \it In a theory of quantum gravity, for any sufficiently large region
in a smooth background spacetime, the entanglement entropy between the degrees
of freedom describing the given region with those describing its complement is
finite and to leading order, takes the form given in eq.~\reef{prop0}.
 \vskip 3pt
\noindent \rm Of course, an implicit assumption here is that the usual
Einstein-Hilbert action (as well as, possibly, a cosmological constant term)
appears as the leading contribution to the low energy effective gravitational
action.
The appearance of this geometric entropy \reef{prop0} in such a general context
emphasizes that in quantum gravity, the description of any macroscopic
spacetime (even flat Minkowski space) entails a state with a great deal of
nontrivial structure in terms of the microscopic degrees of freedom of the
theory.
Further with this proposal, the area law \reef{prop0} for entanglement entropy
serves as a characteristic signature for the emergence of a semiclassical
metric in a theory of quantum gravity. The remainder of the paper will then
present various lines of evidence that support our conjecture. We begin with
some discussion of the relevant background material.

\vspace{.5em}

\noindent {\bf Entanglement entropy:} Entanglement entropy has emerged as a
useful measure of entanglement which can be used to characterize the
correlations in a quantum system 
\cite{wenx}. Given a subsystem $A$ described by 
the reduced density matrix $\rhoa$, the entanglement entropy corresponds to the
von Neumann entropy of the corresponding density matrix: $S_\mt{EE}=-{\rm
tr}\!\left[\,\rhoa \log \rhoa\right]$. In the context of quantum field theory
(QFT), when one considers the entanglement between a region and its
complement,\footnote{
Here and throughout the following, these are spatial regions on a fixed Cauchy
surface.} one finds that the entanglement entropy is UV divergent because of
short range correlations in the vicinity of the `entangling surface' $\Sigma$
separating the two regions. If the calculation is regulated with a short
distance cut-off $\delta$, the leading contributions for a QFT in $d$ spacetime
dimensions generically take the form
 \be
 S_\mt{EE}=c_0\frac{R^{d-2}}{\delta^{d-2}}+c_2\frac{R^{d-4}}{\delta^{d-4}}
 +\cdots\,,
 \label{diverg0}
 \ee
where $R$ is some (macroscopic) scale characterizing the geometry of $\Sigma$.
In fact, a closer examination shows that each of these terms has a precise
geometric interpretation involving 
an integration of various factors over the boundary $\Sigma$ \cite{geob}. For
example, the leading term 
yields the famous `area law' result with $S_\mt{EE}\simeq \tilde{c}_0
\cA_\Sigma/\delta^{d-2}$ \cite{bh8}. Unfortunately the dimensionless
coefficients $c_{k}$ appearing in these power law divergent terms above are
sensitive to the details of the UV regulator.

Entanglement entropy has also been discussed in the context of the AdS/CFT
correspondence and more broadly, of gauge/gravity duality \cite{rt1}. Given a
particular holographic framework, the entanglement entropy in the
$(d-1)$-dimensional boundary theory between a spatial region $A$ and its
complement is calculated by extremizing the following expression
 \be
S(A) = \frac{2\pi}{\lp^{d-2}}\ \mathrel{\mathop {\rm
ext}_{\scriptscriptstyle{\m\sim A}} {}\!\!} \left[\cA(\m)\right]
 \labell{define}
 \ee
over $(d-2)$-dimensional surfaces $\m$ in the bulk spacetime which are
homologous to the boundary region $A$.
In particular then, the boundary of $\m$ matches the `entangling surface'
$\Sigma=\partial A$ in the boundary geometry. Implicitly, eq.~\reef{define}
assumes that the bulk theory is well approximated by classical Einstein
gravity. Hence this expression \reef{define} bears a striking resemblance to
black hole entropy, however, in general, the surfaces $\m$ do not coincide with
an event horizon in the bulk spacetime, or even the boundary of the natural
`causal wedge' associated with the boundary region \cite{wedge}. In the present
context, we present this as the first indication that eq.~\reef{prop0} has a
wide applicability. Further, we might note that there has already been some
speculation that evaluating this expression on other surfaces in the bulk
geometry may yield interesting entropic measures of entanglement in the
boundary theory \cite{wedge,mark}.

\vspace{.5em}

\noindent {\bf S$_{\rm \bf BH}$ = S$_{\rm \bf EE}$:} In fact, the initial
proposals \cite{bh8} to consider what is now called entanglement entropy
stemmed from attempts to understand the entropy of black holes. The observation
is that (a cross-section of) the horizon plays the role of an entangling
surface separating the degrees of freedom between the interior and exterior of
the black hole. Hence it was suggested that quantum correlations between these
two regions might account for the black hole entropy. Recall that in evaluating
the entanglement entropy for a field theory, the leading term obeys the desired
area law:
 \beq
S_\mt{EE}=c_0 \frac{\cA_\Sigma}{\delta^{d-2}}+\cdots\,. \labell{simp}
 \eeq
Of course, this is suggestive. In particular, if one had $\delta\simeq\lp$ in a
quantum theory of gravity, then eq.~\reef{simp} could reproduce
eq.~\reef{prop0}. While it is natural to think that gravity should regulate the
entanglement entropy, the precise mechanism is unclear \cite{ted0}.
Furthermore, as was commented above, these UV sensitive terms in the
entanglement entropy are not universal. That is, different choices of regulator
will yield different values for the dimensionless coefficient $c_0$.

The latter issue was partially resolved in \cite{suss}. There the suggestion
was that this area law contribution of `low energy' degrees of freedom actually
renormalizes the bare area term: $S_0 =\cA/4G_0$ where $G_0$ is the `bare'
Newton's constant,\footnote{Here and in the following, we set $\hbar=1=c$.} in
the sense of perturbative QFT. That is, the renormalization of the Einstein
term in the effective action coming from integrating out quantum fields also
yields a power law divergence of the form
 \be
\Delta\!\left(\frac1G\right)= \frac{{\tilde c}_0}{\delta^{d-2}}\ .
\labell{renorm}
 \ee
Here again, the dimensionless coefficient ${\tilde c}_0$ is also regulator
dependent. However, the proposal of Susskind and Uglum \cite{suss} is that for
a given choice of regulator: $c_0={\tilde c}_0/4$. Hence, the area law term in
the entanglement entropy across the black hole horizon is precisely $S_\mt{EE}=
\frac{\cA}4 \,\Delta\!\left( \frac1{G} \right)$. While there was a great deal
of work done to confirm this idea in explicit examples \cite{revue,exam},
unfortunately, there were cases where it appeared that the desired matching was
not achieved. However, this confusion was recently resolved in \cite{luty}. The
general arguments there confirm 
the proposal of \cite{suss} for any fields
with spin $0$, $1/2$, $1$ and/or $3/2$
to all orders in the perturbation expansion of the corresponding relativistic
QFT in
any 
static curved spacetime background with a (bifurcate) Killing horizon in any
dimension. Unfortunately, there are still technical issues in extending these
arguments to spin-2 fields, \ie the graviton itself. For later purposes, we
note that these results confirm that the area term in $S_\mt{EE}$ can be seen
as renormalizing the geometric entropy \reef{prop0} for a Rindler horizon in
flat space.

We also refer the interested reader to another recent paper \cite{ted}, where a
related result was presented. The perspective advanced there was that as the
renormalization scale is varied, one is simply trading contributions of the
horizon entropy between the bare geometric term and the entanglement entropy of
the low energy degrees of freedom.

While it seems then that the state of affairs with regards to \cite{suss} has
been clarified, the situation is still widely regarded as unsatisfactory. The
horizon entropy now takes the form:
 \be
S_{BH}=S_0+S_\mt{EE}=\frac{\cal A}{4G_0}+ \frac{\cal
A}{4}\,\Delta\!\left(\frac1G \right)+\cdots
 \labell{satisfy}
 \ee
This leaves the question of how to interpret the bare term in the above
contribution. One suggestion was to consider 
models of `induced gravity' \cite{sak} in which there is no bare Einstein term,
\ie $1/G_0=0$. Rather the entire gravity action is generated by quantum
fluctuations of the other degrees of freedom, \eg some heavy fields. Hence in
such a scenario, the entire horizon entropy is also accounted for by the
entanglement entropy of the latter degrees of freedom \cite{induce}. In fact,
we would observe that most modern approaches to quantum gravity take
essentially this perspective. While there may be some fundamental description
of the theory in terms of microscopic degrees of freedom, graviton excitations,
the low energy Einstein action and the spacetime geometry itself typically
arise as some collective or emergent phenomena. With this outlook, it is
natural then to view the metric as an effective 
macroscopic variable \cite{magic} and further then the `off-shell' method
\cite{important} to calculating horizon entropy is precisely a calculation of
the entanglement entropy in terms of these macroscopic variables.\footnote{Let
us note that the `off-shell' method can be extended to higher curvature
theories of gravity, in which case it reproduces precisely the Wald entropy.
See the discussion around footnote 15 in \cite{cthem}.}

It should be noted that, in low-energy processes, the change $\Delta S_\mt{EE}$
in the entanglement entropy reproduces the Bekenstein-Hawking formula
\cite{Bianchi:2012br},
\begin{equation}
\Delta S_\mt{EE}=\frac{\Delta \mathcal{A}}{4 G}\,,
\end{equation}
with $G$ being the low-energy Newton constant, and $\Delta \mathcal{A}$ the
change in area of the event horizon. The variation $\Delta S_\mt{EE}$ is finite
and insensitive to the UV physics because a low-energy process affects only the
IR part of the entropy. Moreover it is universal because of the universal
coupling of gravitons to the energy-momentum tensor. This result from
perturbative quantum gravity provides further support to the idea that the
entropy $S_{BH}$ is due to entanglement as in eq.~(\ref{satisfy}).

Of course, it is reassuring to confirm the expectation that $S_{BH}=S_\mt{EE}$
with explicit microscopic models. Here one classic example is the eternal AdS
black hole, in the AdS/CFT correspondence, where the horizon entropy is
interpreted in terms of the entanglement entropy between the microscopic
degrees of freedom of the boundary CFT and its thermofield double \cite{juan}.
In fact, a similar interpretation extends quite generally to asymptotically AdS
spacetimes with a Killing horizon, \eg \cite{cthem,mark2}. Similarly this area
law arises from calculations in loop quantum gravity and spin foams, where the
entropy of the horizon at the leading order is given by the entanglement
entropy of spin-network links crossing the entangling surface
\cite{bianchi:ilqgs2012,Donnelly:2011hn}. Furthermore, various simplified
models of induced gravity also illustrated this point
\cite{induce,rob06,fur06}.

\vspace{.5em}

\noindent {\bf Entanglement Hamiltonians:} Recall that the first step in
calculating $S_\mt{EE}$ for some region $A$ was to calculate the density matrix
$\rhoa$. Here we note that an essential role played by $\rhoa$ is to encode the
correlators of operators whose support lies in $A$. In fact, by causality,
$\rhoa$ describes such physics throughout the causal development $\cal D$ of
$A$.\footnote{The causal development of $A$ is the set of all points $p$ for
which all causal curves through $p$ necessarily intersect $A$.} Now since the
reduced density matrix is both hermitian and positive semidefinite, it can be
expressed as
 \be
\rhoa=e^{-\ha} \labell{important}
 \ee
for some hermitian operator $\ha$. In the literature on axiomatic quantum field
theory, $\ha$ is known as the `modular Hamiltonian' \cite{haag} while the same
operator is referred to as the `entanglement Hamiltonian' in the condensed
matter literature \cite{cmt}. We note that $\ha$ and the unitary operator
$U(s)=\rho^{i s}=e^{-i \ha s}$ play an important role in axiomatic approaches
to establish various formal properties of $\rhoa$ and the algebra of operators
on $\cal D$. However, we must emphasize that generically $\ha$ is not a local
operator and $U(s)$ does not generate a local (geometric) flow on $\cal D$. For
example, if we begin with a local operator defined at a point, ${\cal
O}=\phi(x)$, then generally the operator ${\cal O}(s)= U(s) {\cal O}\,
U^\dagger(s)$ becomes an operator with support over an extended region within
$\cal D$. To be explicit, we might expect $\ha$ to have 
the schematic form
 \begin{align}
 \ha=&\int d^{d-1}x\, \gamma^{\mu\nu}_1(x)\,T_{\mu\nu} +\label{beast}\\
 &\!\!+\int d^{d-1}x \int
d^{d-1}y\,  \gamma^{\mu\nu;\rho\sigma}_2(x,y)
\,T_{\mu\nu}T_{\rho\sigma}+\cdots\,.\nonumber
 \end{align}
Furthermore, in general, we should expect that other operators beyond the
stress tensor can also appear in this expansion.  Of course, there are special
cases where the modular flow and the modular Hamiltonian are in fact local. A
well-known example is given by the Minkowski vacuum state restricted to the
Rindler wedge $\R$. That is, taking the entangling surface to be the line
$\Sigma=\lbrace t=0,x=0\rbrace$ and the region $A$, the half-space $x>0$ (and
$t=0$), then the corresponding causal development ${\cal D}\equiv\R$ is a wedge
of Minkwoski space. In this case for {\it any QFT}, the modular Hamiltonian is
just
 \be
\ha= 2\pi K +c' = -2\pi \int_{x>0} \!\!\!\!\!\!d^{d-2}y\,dx\,\left[x\,
T_{00}\right] + c'
 \labell{bang}
 \ee
where $K$ 
is the boost generator in the $x$ direction and $c'$ is a constant introduced
to ensure the unit trace of the density matrix. This result is commonly known
as the Bisognano-Wichmann theorem \cite{bisognano}. In this special case, the
operator $U(s)$ translates operators along the boost orbits within $\R$.
Further, of course, interpreted in the sense of Unruh \cite{Unruh},
eq.~\reef{important} describes a thermal state with respect to this notion of
`time' translations.

\begin{figure}[t]
\begin{tabular}{r}
\includegraphics[width=.4\textwidth]{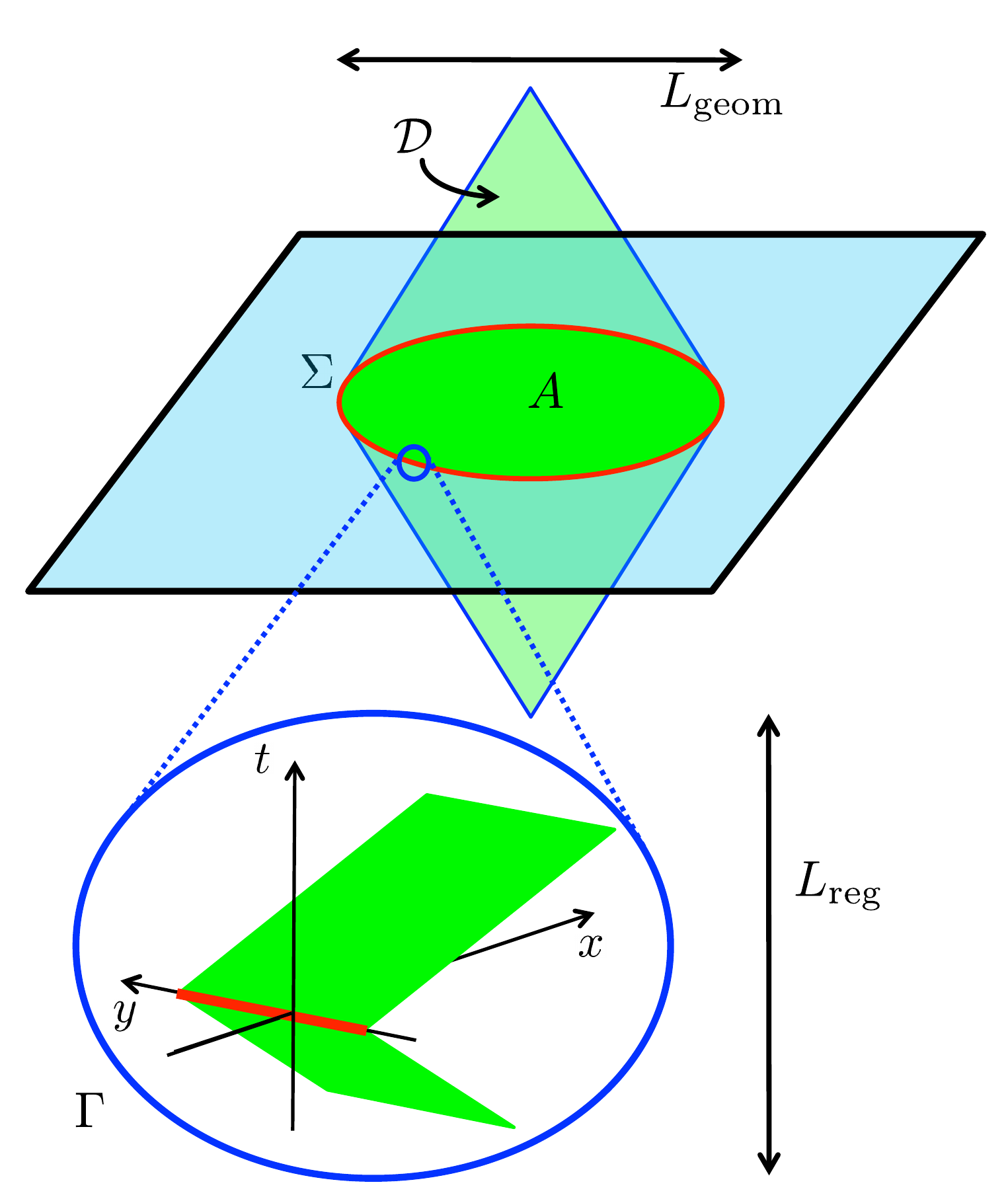}\\
\end{tabular}
\caption{(Colour Online) Consider a region $A$ on a Cauchy slice of some
smooth background
geometry. The density matrix $\rhoa$ controls the physics throughout the
causal domain $\cal D$. This geometry varies only on some large distance
scales $L_{\text{geom}}$. We consider a spacetime region $\Gamma$ of size
$L_{\text{reg}}\ll L_{\text{geom}}$ near the entangling surface $\Sigma$.
Within this region, the spacetime looks
like flat space and the light sheets defining $\partial\cal D$ look like
Rindler horizons.} \label{picture}
\end{figure}
Now given this formalism, let us turn to the question of calculating the
entanglement entropy of some general region. We wish to frame the discussion in
the context of a 
general curved background spacetime in which the curvatures are slowly varying.
Within this background, we choose some smooth Cauchy surface and on this
surface a smooth entangling surface $\Sigma$. We imagine that we are evaluating
the entanglement entropy for a collection of quantum fields in this framework.
The QFT will be provided with some regulator that introduces a short distance
cut-off $\delta$. The latter is much smaller than any geometric scales
$L_{\text{geom}}$ that arise in defining the background, the Cauchy surface and
the entangling surface, \ie $\delta\ll L_{\text{geom}}$. Now let us zoom in on
a small spacetime region $\Gamma$ of size $L_{\text{reg}}\ll L_{\text{geom}}$
near the entangling surface $\Sigma$, as illustrated in figure \ref{picture}.
We imagine that the problem is such that we can choose $\delta\ll
L_{\text{reg}} \ll L_{\text{geom}}$. Hence within $\Gamma$, the spacetime looks
like flat space, the portion of the entangling surface looks simply like a
straight line, and the light sheets defining $\partial\cal D$ are flat surfaces
extending from this line like Rindler horizons.

Now given the general setting in which we have framed our discussion, it would
be difficulat to consider details of the vacuum state for the QFT without
further information. In fact, it may well not be possible to define a unique
vacuum state in general. However, rather than focussing on such a precise
state, we instead consider general states with the property that correlators in
these states reproduce the standard UV singularities of the Minkowski vacuum.
For free fields, this is essentially the definition of Hadamard states, and for
interacting fields, this Hadamard-like property can be seen as a defining
characteristic of the relevant states \cite{Hollands:2008vx}. For any such
states, this ensures that within the region $\Gamma$, any correlators have the
same form as in Minkowski space up to small corrections. Hence the short
distance part of $\rhoa$ that encodes these correlators must have the same
structure as in flat space. But then as discussed above, we know precisely the
form of this density matrix. In particular, expressing $\rhoa$ as in
eq.~\reef{important}, the leading order contribution to the entanglement
Hamiltonian must be precisely the Rindler Hamiltonian, \ie $\ha= 2\pi K +
\cdots$ as in eq.~\reef{bang}.\footnote{As an aside, let us note that this
result agrees with the usual intuition about 
the local character of UV physics. In particular, as stressed above, the full
entanglement Hamiltonian will generally be a nonlocal operator but here we have
argued that the leading UV sensitive contribution to $S_\mt{EE}$ is controlled
by a local term in $\ha$. The same intuition suggests that it is natural to
think that the other UV divergent contributions to $S_\mt{EE}$ should also be
governed by local terms in $\ha$ \cite{new1}.}

Now this conclusion has two interesting implications. First, we know that the
`density' of the entanglement entropy for a Rindler horizon yields a constant
leading divergence, \ie $s\simeq\frac{c_0}{\delta^{d-2}}+\cdots$. While the
constant $c_0$ depends on our precise choice of regulator, in the present
framework the regulator is fixed and 
so each element of the entangling surface makes precisely the same
contribution. Hence integrating over the entangling surface, we will find the
leading singularity will be a contribution of the form $S_\mt{EE}\simeq
c_0\frac{\cA_\Sigma}{ \delta^{d-2}} +\cdots$. That is, for any calculation of
$S_\mt{EE}$ within the general context set out above, we will find that the
leading contribution to the entanglement entropy is an area law term.
Furthermore, at this point, we may invoke the results of \cite{luty} 
as applied to Rindler horizons to further relate the constant $c_0$ to the
renormalization of the effective Newton's constant in this theory. Hence we
arrive at the conclusion that the leading contribution to the entanglement
entropy takes the form
 \be
S_\mt{EE}\simeq \frac{\cA_\Sigma}{4}\ \Delta\!\left(\!\frac1G\!\right)
+\cdots\,.
 \labell{bang2}
 \ee
That is, for any general region in a smooth background spacetime, as described
above, the entanglement entropy of the low energy fields has precisely the
necessary form to renormalize the expression given in eq.~\reef{prop0}. At the
very least, this is a highly nontrivial consistency check of the proposal made
at the outset of this paper. 
That is, in order for
eq.~\reef{prop0} to be consistent with the coupling of the perturbative quantum
fields to gravity, these low energy degrees of freedom must contribute to the
entanglement entropy precisely as in eq.~\reef{bang2} for general regions.

Of course, as with black hole entropy, this discussion leaves open the question
of how to interpret the bare area term in $S_\mt{EE}$, which our proposal
requires to be present. Here, we would advocate that applying the `off-shell'
method to calculate the Rindler entropy within each region indicates that this
bare term must be present as the entanglement entropy of the microscopic
gravity degrees of freedom. There will also be many higher order corrections,
\eg corresponding to integrals of both intrinsic and extrinsic curvatures over
the entangling surface. It is not clear that all of these will be associated to
the renormalization of various gravitational couplings \cite{new1,new2}.

Of course, as before, it is much more satisfying to explicitly realize 
the desired result with calculations within a given microscopic model. Here one
can draw evidence from two sources: First, the Randall-Sundrum II braneworld
\cite{RS2} provides an example of an induced gravity model. In particular, it
has been observed that in this framework, using holographic prescription
for entanglement entropy \cite{rt1}, black hole entropy corresponds to
entanglement entropy \cite{rob06}. However, it is also straightforward to show
that the area term in eq.~\reef{prop0} appears for any sufficiently large
region, irrespective of whether or not the entangling surface corresponds to an
event horizon \cite{rob06,new2}. Similar results were noted in \cite{fur06} for
other simple induced gravity models using heat kernel techniques. Finally
we turn to loop quantum gravity to find support for our conjecture.\\

\noindent {\bf Spin Foam Models:} In `loop quantum gravity', a smooth
macroscopic geometry is expected to emerge from a description of space and
spacetime which is discrete at a fundamental level \cite{LQG}. There has been
recent progress in the understanding of black hole entropy in this context
\cite{bianchi:ilqgs2012,Ghosh:2011fc} and so it is natural to ask whether these
models give some evidence for our conjecture 
that general regions of macroscopic spacetimes carry an entanglement entropy
given by eq.~\reef{prop0}.

Consider a cellular decomposition of a three-dimensional manifold, for
instance, a triangulation. A spin-network graph with a node in each cell
and a link connecting nodes in neighbouring cells is said to be dual to this triangulation.
Lorentz-group representations label the links of the graph and determine a quantum
geometry of the triangulation. Generically such states are highly entangled \cite{Donnelly:2011hn}. In
particular, we consider the vacuum state defined using the covariant spinfoam
dynamics, which has the properties that it is invariant under local Lorentz
transformation and time translations. Now, even neglecting interactions between
different links, the state has entanglement associated to the endpoints of each
link. In the cellular picture, the quantum geometries of two nearby cells in
the three-dimensional manifold are entangled.

Now we consider a three-dimensional region $A$ in the manifold. The cellular
decomposition induces on the boundary $\Sigma$ of the region a tessellation in
two-dimensional cells. In the dual picture these are links $l$ crossing the
surface $\Sigma$. Exactly as discussed above, the relevant part of the reduced
density matrix $\rhoa$ can be written in the form (\ref{important}) with the
entanglement Hamiltonian
\begin{equation}\textstyle
\ha=2\pi \sum_l K_l\;+\,\log Z\,.
\end{equation}
The sum is over the links $l$ that cross the entangling surface $\Sigma$, and
$K_l$ is the hermitian generator of boosts in the unitary representation of the
Lorentz group associated to the link. This expression has the same form of
eq.~(\ref{bang}) for the QFT case. The term $\log Z$ provides the normalization
of the density matrix $\rhoa=e^{-\ha}$, 
\ie this term provides the constant $c'$ in eq.~\reef{bang}. The entanglement
entropy is 
now easily computed:
\begin{equation}\textstyle
S_\mt{EE}=-\text{Tr}(\rhoa \log \rhoa)\,=\,2\pi\text{Tr}(\sum_l K_l\;\rhoa)\,+\,\log Z\,.
\end{equation}
The simplicity constraint on representations of the Lorentz group allows 
us to express the first term as the area $\mathcal{A}_\Sigma$ of the entangling
surface \cite{bianchi:ilqgs2012}. The second term is proportional to the number
$\mathcal{N}$ of links crossing $\Sigma$, so that we have
\begin{equation}
S_\mt{EE}=\, \frac{\mathcal{A}_\Sigma}{4 G_0 }\;+\;\mu(\gamma)\ 
\mathcal{N}\,,
\end{equation}
where $\mu$ is a chemical potential that depends on the Immirzi parameter
$\gamma$ \cite{Ghosh:2011fc}. The entanglement entropy is finite because the
theory has no degrees of freedom below the scale $\ell_{LQG}=(8\pi\gamma\,
G_0)^{1/2}$, the physical cut-off scale in loop quantum gravity.  As the area
$\mathcal{A}_\Sigma$ is proportional to $\mathcal{N}$, the second term can be
understood as a finite renormalization of $G_0$ and be reabsorbed in the first
term in the same way as described in eq.~(\ref{satisfy}), thus providing
further evidence for our conjecture.\\

\noindent {\bf Discussion:} We have proposed that the Bekenstein-Hawking
formula has a much wider applicability that previously considered. In fact, our
conjecture is that eq.~\reef{prop0} corresponds to the leading contribution to
the entanglement entropy for any sufficiently large region in a theory of
quantum gravity. Evidence for this conjecture was presented from four directions:\\
\noindent i) In the AdS/CFT correspondence, the well-tested prescription
for holographic entanglement entropy \cite{rt1} clearly assigns an entropy to
large classes of surfaces which are unrelated to horizons, with precisely
eq.~\reef{prop0} as the leading term.\\
\noindent ii) In examining quantum fields in curved spacetime, for any large
region, the leading contribution to the entanglement entropy is an area term
and the coefficient of this term matches precisely the renormalization of
Newton's constant in eq.~\reef{prop0}. Further, applying the `off-shell' method
to calculate the Rindler entropy locally along the entangling surface suggests
the presence of a bare term ${\cal A}/4G_0$, as well.\\
\noindent iii) In simplified models of induced gravity, the leading term to the
entanglement entropy for large regions is finite and takes precisely the
form given by eq.~\reef{prop0} \cite{rob06,fur06,new2}.\\
\noindent iv) Our preliminary investigations of spin foam models indicate that
general regions will carry a finite entanglement entropy, again with the
leading term described by eq.~\reef{prop0}.\\
\noindent We feel that combining these results provides strong evidence for our
conjecture as a general result.

Our proposal 
demands that quantum gravity 
effects two essential features for entanglement entropy: First, it `regulates'
entanglement entropies for general regions. This might be seen as another
realization of the general lore that quantum gravity contains fewer states than
quantum field theory. The second property is that this regulator yields a
simple universal result, \ie eq.~\reef{prop0}. This property would seem to rely
on the universal couplings of the effective Einstein theory emerging at low
energies \cite{Bianchi:2012br}. We expect this universality is a unique feature
of the entanglement entropy. For example, the \ren entropies
\cite{renyi0,karol}, which would provide another measure of the entanglement
between regions, should also exhibit an area law behaviour at leading order.
However, the precise coefficient of the area term would likely depend on the
microscopic details of the underlying quantum gravity theory.

In quantum many-body systems, the entanglement entropy typically satisfies an
area law \cite{Eisert:2008ur}. However, this is not the typical behaviour for
generic states in the full Hilbert space \cite{gen}. Hence it seems that the
locality of the underlying Hamiltonian restricts the entanglement of the
microscopic constituents in the low energy states of these systems. This
feature was central to the recent development of tensor network techniques to
better understand the nature of quantum matter \cite{net}. Drawing an analogy
here, we expect that generic states in the full Hilbert space of quantum
gravity will not correspond to anything resembling a smooth spacetime. Rather
states describing smooth macroscopic spacetimes require a certain structure for
the short-range entanglement such that we get the area law behaviour
\reef{prop0} as conjectured here. Hence this discussion suggests that the area
law entanglement of eq.~\reef{prop0} can be regarded as a signature of states
that approximate smooth spacetimes in quantum gravity. In this way then,
eq.~\reef{prop0} provides us with a glimpse into the quantum architecture of
macroscopic spacetime geometries.

In closing, we consider some future directions: Clearly, it is of interest to
further develop the calculations for the spin foam models. Given the discussion
in the preceding paragraph, it would be interesting if the tensor network
approaches in condensed matter could provide new lessons on how to deal with
discrete models of quantum gravity. It would also be interesting to identify
analogous calculations in the context of string theory, as well as to better
understand our proposal in a holographic framework. Moreover, it would be
useful to identify if this entanglement entropy has an operational meaning.
Certainly, when applied to black hole horizons, there is an interpretation in
terms of thermodynamic entropy of the black hole, where it should also
correspond to a counting of states. Here we might note that low-energy
perturbations of the entanglement entropy 
seem to admit such a
thermodynamic interpretation \cite{Bianchi:2012br,Taka:thermo}.\\


\noindent {\bf Acknowledgments:} We thank Horacio Casini, Borun Chowdhury,
Bartlomiej Czech, Roberto Emparan, Laurent Freidel, Ben Freivogel, Dmitry
Fursaev, Daniel Gottesman, Ted Jacobson, Stefano Liberati, Markus Luty, David
Mateos, Arif Mohd, Razieh Pourhasan, Carlo Rovelli, Misha Smolkin, Rafael
Sorkin, Brian Swingle and Bob Wald for discussions. Research at Perimeter
Institute is supported by the Government of Canada through Industry Canada and
by the Province of Ontario through the Ministry of Research \& Innovation. RCM
also acknowledges support from an NSERC Discovery grant and funding from the
Canadian Institute for Advanced Research. EB also acknowledges support from a
Banting Postdoctoral Fellowship from NSERC.

\end{document}